\documentclass{article} 
\usepackage{latexsym}
\usepackage{graphicx}

\def\gsize{0.95} %グラフのサイズ　<0.74では線が細くなってしまう
\def\gs2{0.3} %グラフのサイズ　<0.74では線が細くなってしまう
\def\g3{0.8} %グラフのサイズ　<0.74では線が細くなってしまう

\title{Statistical Mechanics of Time Domain Ensemble Learning}

\author{
Seiji MIYOSHI
\thanks{
Department of Electronic Engineering,
Kobe City College of Technology,
8--3 Gakuenhigashimachi, Nishi-ku, 
Kobe-shi, 651--2194 Japan
}
\and
Tatsuya UEZU
\thanks{
Graduate School of Humanities and Sciences, 
Nara Women's University, 
Kitauoyahigashi-machi, Nara, 630--8506 Japan
}
\and
Masato OKADA
\thanks{
Division of Transdisciplinary Sciences, 
Graduate School of Frontier Sciences, The University of Tokyo,
5--1--5 Kashiwanoha, Kashiwa-shi, Chiba, 277--8561 Japan
\newline
RIKEN Brain Science Institute,
2--1 Hirosawa, Wako-shi, Saitama, 351--0198 Japan
\newline
JST PRESTO
}
}

\begin{document}
\maketitle

%\large

%**********************************************************
\section{Abstract}
%**********************************************************
Conventional ensemble learning combines students
in the space domain. 
On the other hand, in this paper we combine 
students in the time domain
and call it time domain ensemble learning.
In this paper, we analyze the generalization performance
of time domain ensemble learning
in the framework of online learning 
using a statistical mechanical method.
We treat a model in which
both the teacher and the student
are linear perceptrons with noises.
Time domain
ensemble learning is twice as effective as 
conventional space domain ensemble learning.

\vspace{3mm}

\noindent
{\it Keywords}: 
ensemble learning, 
online learning, 
generalization error,
statistical mechanics

%**********************************************************
\section{Introduction}
%**********************************************************
Learning is to infer the underlying rules that dominate 
data generation using observed data.
Observed data are input-output pairs from a teacher
and are called examples.
Learning can be roughly classified into batch learning and
online learning \cite{Saad}.
In batch learning, given examples are used more than once.
In this paradigm, a student will give the correct answers
after training if that student has adequate degree of freedom.
However, it is necessary to have a long time and 
a large memory in which many examples are stored.
On the contrary, in online learning examples 
used once are then discarded.
In this case, a student cannot give correct answers 
for all the examples used in training.
However, there are merits, for example,
a large memory for storing many examples is not necessary,
and it is possible to follow a time variant teacher. 

Recently, we \cite{Hara,PRE} 
analyzed the generalization performance
of some models
in a framework of online learning
using a statistical mechanical method
\cite{PRE,JPSJ2006a,JPSJ2006b}.
Ensemble learning means 
to combine many rules or learning machines
(called students in this paper) that perform poorly;
it has recently attracted the attention 
of many researchers
\cite{Abe,www.boosting.org,Krogh,Urbanczik,Hara}.
The diversity or variety of students is important
in ensemble learning. 
We showed that
the three well-known rules, 
Hebbian learning, perceptron learning, and
AdaTron learning
have different characteristics in their 
affinities for ensemble learning, that is in 
``maintaining diversity among students"\cite{PRE}．
In that process, the following points
were proven subsidiarily
\cite{IBIS2004,NC200503,Maeda,Uezu}.
The student vector 
doesn't converge in one direction but continues moving
in an unlearnable case\cite{Inoue,Inoue2}.
Therefore, 
we also analyzed the generalization performance
of a student supervised by a moving teacher that goes around
a true teacher\cite{JPSJ2006a}.
As a result, 
it was proven that
the generalization error of a student
can be smaller than
a moving teacher,
even if the student only uses examples 
from the moving teacher.
In an actual human society, a teacher
observed by a student
does not always present the correct answer.
In many cases, the teacher is learning
and continues to change.
Therefore, the analysis of such a model
is interesting for considering the 
analogies between statistical learning theories 
and an actual human society.

In conventional ensemble learning,
the generalization performance is improved by combining
students who have diversities.
On the other hand, a student does not always
converge in one direction but may continue moving 
in an unlearnable model.
Therefore, the generalization performance in such a model 
must be improved by combining the student itself
at different times even if there is only one student
\cite{Maeda,Uezu}.
Conventional ensemble learning combines students
in the space domain. 
On the other hand, we introduce a method of combining 
the students in the time domain;
we call this time domain ensemble learning.
In this paper, we analyze the generalization performance
of the time domain ensemble learning
using a statistical mechanical method.
We treat a model in which
both the teacher and the student
are linear perceptrons\cite{Hara} with noises.
We obtain the order parameters and generalization errors
analytically 
in a framework of online learning 
using a statistical mechanical method.

%**********************************************************
\section{Model}
%**********************************************************
In this paper we consider a teacher and a student.
They are linear perceptrons with the connection weights
$\mbox{\boldmath $B$}$ and $\mbox{\boldmath $J$}^m$, 
respectively.
Here, $m$ denotes the time step.
For simplicity, the connection weights of the teacher
and the student
are simply called the teacher and the student, respectively.
Teacher $\mbox{\boldmath $B$}=\left(B_1,\ldots,B_N\right)$,
student $\mbox{\boldmath $J$}^m=\left(J^m_1,\ldots,J^m_N\right)$,
and input
$\mbox{\boldmath $x$}^m=\left(x^m_1,\ldots,x^m_N\right)$
are $N$ dimensional vectors.
Each component $B_i$ of $\mbox{\boldmath $B$}$
is independently drawn from ${\cal N}(0,1)$ and fixed,
where ${\cal N}(0,1)$ denotes a Gaussian distribution with
a mean of zero and variance unity.
Each component $J_i^0$
of the initial value $\mbox{\boldmath $J$}^0$
of $\mbox{\boldmath $J$}^m$
is independently drawn from ${\cal N}(0,1)$.
The direction cosine between 
$\mbox{\boldmath $J$}^m$ and 
$\mbox{\boldmath $B$}$ is $R^m$
and that between
$\mbox{\boldmath $J$}^m$ and
$\mbox{\boldmath $J$}^{m'}$ is $q^{m,m'}$.
Each component $x^m_i$ of $\mbox{\boldmath $x$}^m$
is drawn from ${\cal N}(0,1/N)$ independently.
Thus,
\begin{eqnarray}
\left\langle B_i\right\rangle &=& 0, \ \ 
\left\langle \left(B_i\right)^2\right\rangle=1, \\
\left\langle J_i^0\right\rangle &=& 0, \ \ 
\left\langle \left(J_i^0\right)^2\right\rangle=1,\\
\left\langle x_i^m\right\rangle &=& 0, \ \ 
\left\langle \left(x_i^m\right)^2\right\rangle=\frac{1}{N}, \\
R^m&\equiv&\frac{\mbox{\boldmath $B$}\cdot\mbox{\boldmath $J$}^m}
{\|\mbox{\boldmath $B$}\|\|\mbox{\boldmath $J$}^m\|},\\
q^{m,m'}&\equiv&\frac{\mbox{\boldmath $J$}^m \cdot \mbox{\boldmath $J$}^{m'}}
{\|\mbox{\boldmath $J$}^m \|\| \mbox{\boldmath $J$}^{m'}\|}, \label{eqn:qdef}
\end{eqnarray}
where $\langle \cdot \rangle$ denotes a mean.

\begin{figure}[htbp]
\vspace{3mm}
\begin{center}
\includegraphics[width=\gs2\linewidth,keepaspectratio]{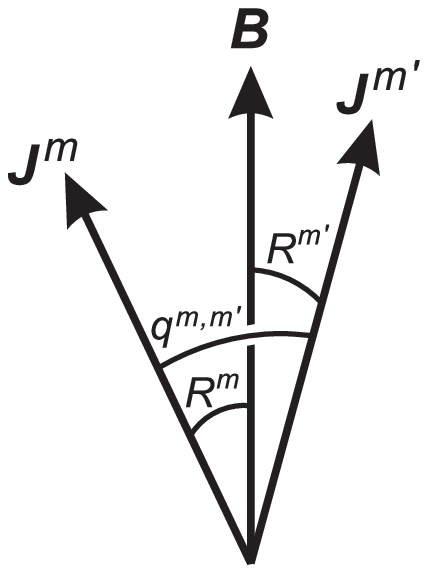}
%-----------------------------------------------
\caption{
Teacher $\mbox{\boldmath $B$}$ and
students $\mbox{\boldmath $J$}^m$
and $\mbox{\boldmath $J$}^{m'}$.
$R^m, R^{m'}$, and $q^{m,m'}$ are direction cosines.
}
%-----------------------------------------------
\label{fig:BJJ}
\end{center}
\end{figure}

Figure \ref{fig:BJJ} illustrates
the relationship among teacher
$\mbox{\boldmath $B$}$,
students $\mbox{\boldmath $J$}^m$ and 
$\mbox{\boldmath $J$}^{m'}$
and the direction cosines
$R^m, R^{m'}$, and $q^{m,m'}$.

In this paper, the thermodynamic limit $N\rightarrow \infty$
is also treated. Therefore,
\begin{equation}
\|\mbox{\boldmath $B$}\|=\sqrt{N},\ \ 
\|\mbox{\boldmath $J$}^0\|=\sqrt{N},\ \ 
\|\mbox{\boldmath $x$}^m\|=1.
\label{eqn:xBJ}
\end{equation}
Generally, the norm $\|\mbox{\boldmath $J$}^m\|$
of the student
changes as the time step proceeds.
Therefore, the ratios $l^m$ of the norm to $\sqrt{N}$
are introduced and are called the length of 
the student. That is,
$\|\mbox{\boldmath $J$}^m\|=l^m\sqrt{N}$.

Both the teacher and the student are linear perceptrons.
Their outputs are
$v^m+n_B^m$ and $u^ml^m+n_J^m$, respectively.
Here,
\begin{eqnarray}
v^m &=& 
 \mbox{\boldmath $B$}\cdot \mbox{\boldmath $x$}^m, \label{eqn:v}\\
u^m l^m &=& 
 \mbox{\boldmath $J$}^m\cdot \mbox{\boldmath $x$}^m, \label{eqn:u}\\
n_B^m &\sim& {\cal N}\left(0,\sigma_B^2\right),\label{eqn:nB}\\
n_J^m &\sim& {\cal N}\left(0,\sigma_J^2\right),\label{eqn:nJ}
\end{eqnarray}
where
${\cal N}(0,\sigma^2)$ denotes a Gaussian distribution with
a mean of zero and variance $\sigma^2$.
That is,
the outputs of the teacher and the student
include independent Gaussian noises with variances of 
$\sigma_{B}^2$ and $\sigma_{J}^2$, respectively.
Then, $v^m$ and $u^m$
obey Gaussian distributions with a mean of zero and 
variance unity.

Let us define the error $\epsilon^m_S$ between 
the teacher $\mbox{\boldmath $B$}$
and the student $\mbox{\boldmath $J$}^m$ alone
by the squared error of their outputs:
\begin{equation}
\epsilon^m_S \equiv 
\frac{1}{2}\left( v^m+n_B^m-u^ml^m-n_J^m\right)^2.
\label{eqn:eS}
\end{equation}

Student $\mbox{\boldmath $J$}^m$
adopts the gradient method as a learning rule
and uses
input $\mbox{\boldmath $x$}$
and an output
of teacher
$\mbox{\boldmath $B$}$
for updates.
That is,
\begin{eqnarray}
\mbox{\boldmath $J$}^{m+1}
&=& \mbox{\boldmath $J$}^{m} 
   -\eta \frac{\partial \epsilon^m_S}{\partial \mbox{\boldmath $J$}^{m}}\\
&=& \mbox{\boldmath $J$}^{m} 
   +\eta \left( v^m+n_B^m-u^ml^m-n_J^m\right)
   \mbox{\boldmath $x$}^{m}, \label{eqn:Jupdate}
\end{eqnarray}
where $\eta$ denotes the learning rate of the student
and is a constant number.
Generalizing the learning rule, 
eq.(\ref{eqn:Jupdate})
can be expressed as
\begin{eqnarray}
\mbox{\boldmath $J$}^{m+1}
&=& \mbox{\boldmath $J$}^{m}
 +f^m
 \mbox{\boldmath $x$}^{m}, \label{eqn:Jupdate-g}
\end{eqnarray}
where $f$ denotes a function
that represents the update amount and 
is determined by the learning rule.

%**********************************************************
\section{Theory}
%**********************************************************
%++++++++++++++++++++++++++++++++++++++++++++++++++++++++++
\subsection{Generalization error}
%++++++++++++++++++++++++++++++++++++++++++++++++++++++++++
Ensemble learning means 
to improve performance
by combining many students that perform poorly.
On the other hand, we use 
just one student and combine its copies
(hereafter called 'brothers')
at different time steps in this paper.
Conventional ensemble learning combines students
in the space domain; on the other hand, we 
combine students in the time domain.
In this paper $K$ brothers
$\mbox{\boldmath $J$}^{m_1}, \mbox{\boldmath $J$}^{m_2},\ldots,
\mbox{\boldmath $J$}^{m_K}$
are combined.
Here, $m_1 \leq m_2 \leq \ldots \leq m_K$.
We use the squared error $\epsilon$
for new input $\mbox{\boldmath $x$}$.
Here, it is assumed that
the Gaussian noises of eqs.(\ref{eqn:nB}) and (\ref{eqn:nJ})
are independently added to 
the teacher and each brother of the ensemble,
respectively.
The weight of each brother $\mbox{\boldmath $J$}^{m_k}$ 
of the ensemble satisfies $C_k>0$.
That is, the error of the ensemble is
%\begin{equation}
%\sum_{k=1}^K C_k 
%\left(\mbox{\boldmath $J$}^{m_k}\cdot\mbox{\boldmath $x$}+n_k \right)
%\label{eqn:output}
%\end{equation}
%and
\begin{equation}
\epsilon=\frac{1}{2}
\left(\mbox{\boldmath $B$}\cdot\mbox{\boldmath $x$}+n_B
-\sum_{k=1}^K C_k 
\left(\mbox{\boldmath $J$}^{m_k}\cdot\mbox{\boldmath $x$}+n_k \right)
\right)^2.
\label{eqn:e}
\end{equation}
%respectively. 
Here,
$n_B \sim {\cal N}\left(0,\sigma_B^2\right)$ and
$n_k \sim {\cal N}\left(0,\sigma_J^2\right)$.

A goal of statistical learning theory
is to theoretically obtain generalization errors.
Since a generalization error is the mean of errors 
over the distribution of the new input $\mbox{\boldmath $x$}$
and noises $n_B, n_k, k=1,\ldots,K$,
the generalization error $\epsilon_g$ of the ensemble is
calculated as follows:
\begin{eqnarray}
\epsilon_g
&=& \int d\mbox{\boldmath $x$}dn_B \left(\prod_{k=1}^K dn_k\right)
         p(\mbox{\boldmath $x$})p(n_B)
         \left(\prod_{k=1}^K p(n_k)\right) \epsilon \\
&=& \int dv \left(\prod_{k=1}^K du_k\right) dn_B 
          \left(\prod_{k=1}^K dn_k \right)
          p(v,\{u_k\})p(n_B)
          \left(\prod_{k=1}^K p(n_k)\right)
          \frac{1}{2}\left(v+n_B-\sum_{k=1}^K C_k 
          \left(u_k l^{m_k}+n_k \right)\right)^2 \nonumber \\
& & \\
&=& \frac{1}{2}\left(1-2\sum_{k=1}^K C_k l^{m_k}R^{m_k} 
       + 2\sum_{k=1}^K \sum_{k'>k}^K C_k C_{k'} l^{m_k}l^{m_{k'}}q^{m_k,m_{k'}}
       + \sum_{k=1}^K C_k^2(l^{m_k})^2 +\sigma_B^2
       + \sum_{k=1}^K C_k^2 \sigma_J^2 \right), \nonumber \\
& & \label{eqn:eg}
\end{eqnarray}
where
$v=\mbox{\boldmath $B$}\cdot\mbox{\boldmath $x$}$
and
$u_k l^{m_k}=\mbox{\boldmath $J$}^{m_k}\cdot\mbox{\boldmath $x$}$.
We executed integration using the following:
$v$ and $u_k$ obey ${\cal N}(0,1)$.
The covariance between $v$ and $u_k$ is $R^{m_k}$,
that between $u_k$ and $u_{k'}$ is $q^{m_k,m_{k'}}$.
$n_B$ and $n_k$
are independent from other probabilistic variables.

%++++++++++++++++++++++++++++++++++++++++++++++++++++++++++
\subsection{Differential equations for order parameters and
their analytical solutions}
%++++++++++++++++++++++++++++++++++++++++++++++++++++++++++
In this paper, we examine the thermodynamic limit $N\rightarrow \infty$.
Therefore,
updates for (\ref{eqn:Jupdate}) or (\ref{eqn:Jupdate-g})
must be executed $O(N)$ times for the order 
parameters $l, R$, and $q$ to change by $O(1)$.
Thus, the continuous times $t_1,\ldots,t_K$,
which are the time steps $m_1,\ldots,m_K$ normalized
by the dimension $N$,
are introduced as the superscripts 
that stand for the learning process.
To simplify the analysis, we introduced the following auxiliary 
order parameters:
\begin{eqnarray}
r^t &\equiv& l^t R^t, \label{eqn:r} \\
Q^{t,t'} &\equiv& l^t l^{t'} q^{t,t'}. \label{eqn:Q}
\end{eqnarray}

The simultaneous differential equations
in deterministic forms \cite{NishimoriE}, 
which describe the dynamical behaviors of order parameters,
have been obtained based on the self-averaging
of thermodynamic limits as follows:
\begin{eqnarray}
\frac{dl^t}{dt} &=& \langle f^t u^t\rangle 
                   + \frac{\langle (f^t)^2\rangle}{2l^t},
\label{eqn:dldt3} \\
\frac{dr^t}{dt} &=& \langle f^t v^t\rangle, \label{eqn:drdt3} \\
\frac{dQ^{t,t'}}{dt'} 
&=& l^t \langle f^{t'}\bar{u}^{t}\rangle, \label{eqn:dQdt3} 
\end{eqnarray}
where
$t'\geq t$ and
$\bar{u}^t=\mbox{\boldmath $x$}^{t'} \cdot \mbox{\boldmath $J$}^{t}/l^{t}
\sim {\cal N}(0,1)$.

Since linear perceptrons are used in this paper,
the sample averages that appear in the above 
equations can be easily calculated as follows:
\begin{eqnarray}
\langle f^tu^t \rangle &=& \eta (r^t/l^t-l^t), \label{eqn:fu}\\
\langle f^tv^t \rangle &=& \eta (1-r^t), \label{eqn:fv}\\
\langle (f^t)^2 \rangle &=& 
\eta^2 (1+\sigma_B^2 + \sigma_J^2 +(l^t)^2 -2r^t), \label{eqn:f2}\\
\langle f^{t'} \bar{u}^{t} \rangle 
&=& \eta \left(r^{t}-Q^{t,t'}\right)/l^t. \label{eqn:fbarut}
\end{eqnarray}

Since all components 
of teacher $\mbox{\boldmath $B$}$
and the initial student $\mbox{\boldmath $J$}^0$
are independently drawn from ${\cal N}(0,1)$
and because the thermodynamic limit $N\rightarrow \infty$
is also used,
they are orthogonal to each other in the initial state.
That is,
\begin{equation}
R^0=0.
\label{eqn:Rinit}
\end{equation}

In addition,
\begin{equation}
l^0=1
\label{eqn:linit}
\end{equation}
and
\begin{equation}
Q^{t,t}=(l^t)^2
\label{eqn:Qinit}
\end{equation}
using eqs.(\ref{eqn:qdef}) and (\ref{eqn:Q}).
Using these initial conditions,
we can analytically solve the 
simultaneous differential equations (\ref{eqn:dldt3})--(\ref{eqn:fbarut})
as follows:
\begin{eqnarray}
r^t&=& 1-e^{-\eta t}, \label{eqn:rsol} \\
(l^t)^2 &=& 
1+\frac{\eta}{2-\eta}\left(\sigma_B^2+\sigma_J^2\right)
-2e^{-\eta t}
+\left(2-\frac{\eta}{2-\eta}\left(\sigma_B^2+\sigma_J^2\right)\right)
e^{\eta(\eta-2)t}, \label{eqn:l2sol}\\
Q^{t,t'} &=& 1-e^{-\eta t}
+e^{-\eta t'}
+\left((l^t)^2-1\right)e^{-\eta (t'-t)}.
\label{eqn:Qsol}
\end{eqnarray}

Substituting eqs.(\ref{eqn:rsol})--(\ref{eqn:Qsol}) 
into eq.(\ref{eqn:eg}),
the generalization error $\epsilon_g$
can be analytically obtained as a function of
time $t_k,\ k=1,\ldots,K$ as follows:
\begin{eqnarray}
\epsilon_g
&=& \frac{1}{2}\left[1-2\sum_{k=1}^K C_k \left(1-e^{-\eta {t_k}}\right)\right. 
             \nonumber \\
&+& 2\sum_{k=1}^K \sum_{k'>k}^K C_k C_{k'} 
%\nonumber \\
%&\times& 
\left(1-e^{-\eta {t_k}}+e^{-\eta {t_{k'}}}+
\left(\bar{\sigma}^2
-2e^{-\eta {t_k}}
+\left(2-\bar{\sigma}^2\right)
e^{\eta(\eta-2){t_k}}\right)e^{-\eta ({t_{k'}}-{t_k})} \right) \nonumber \\
&+& \left. \sum_{k=1}^K C_k^2
\left(
1+\bar{\sigma}^2
-2e^{-\eta {t_k}}
+\left(2-\bar{\sigma}^2\right)
e^{\eta(\eta-2){t_k}}
\right)
+\sigma_B^2+\sum_{k=1}^K C_k^2 \sigma_J^2
\right],\label{eqn:eg2}\\
\bar{\sigma}^2&=&\frac{\eta}{2-\eta}\left(\sigma_B^2+\sigma_J^2\right).
\label{eqn:sbar}
\end{eqnarray}

%**********************************************************
\section{Results and Discussion}
\begin{figure}[htbp]
\begin{minipage}{.500\linewidth}
\begin{center}
\includegraphics[width=\gsize\linewidth,keepaspectratio]{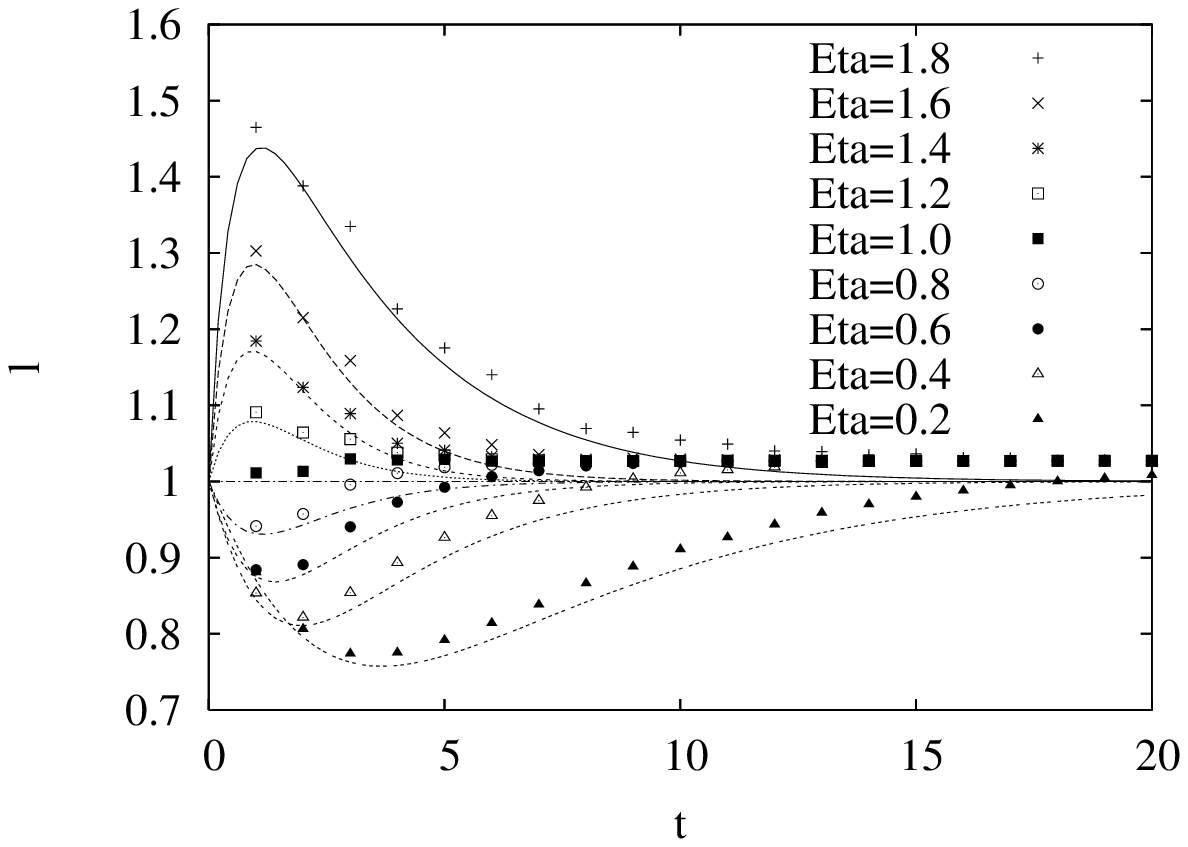}\\
(a) Dynamical behaviors of $l^t$.
\end{center}
\end{minipage}
\begin{minipage}{.500\linewidth}
\begin{center}
\includegraphics[width=\gsize\linewidth,keepaspectratio]{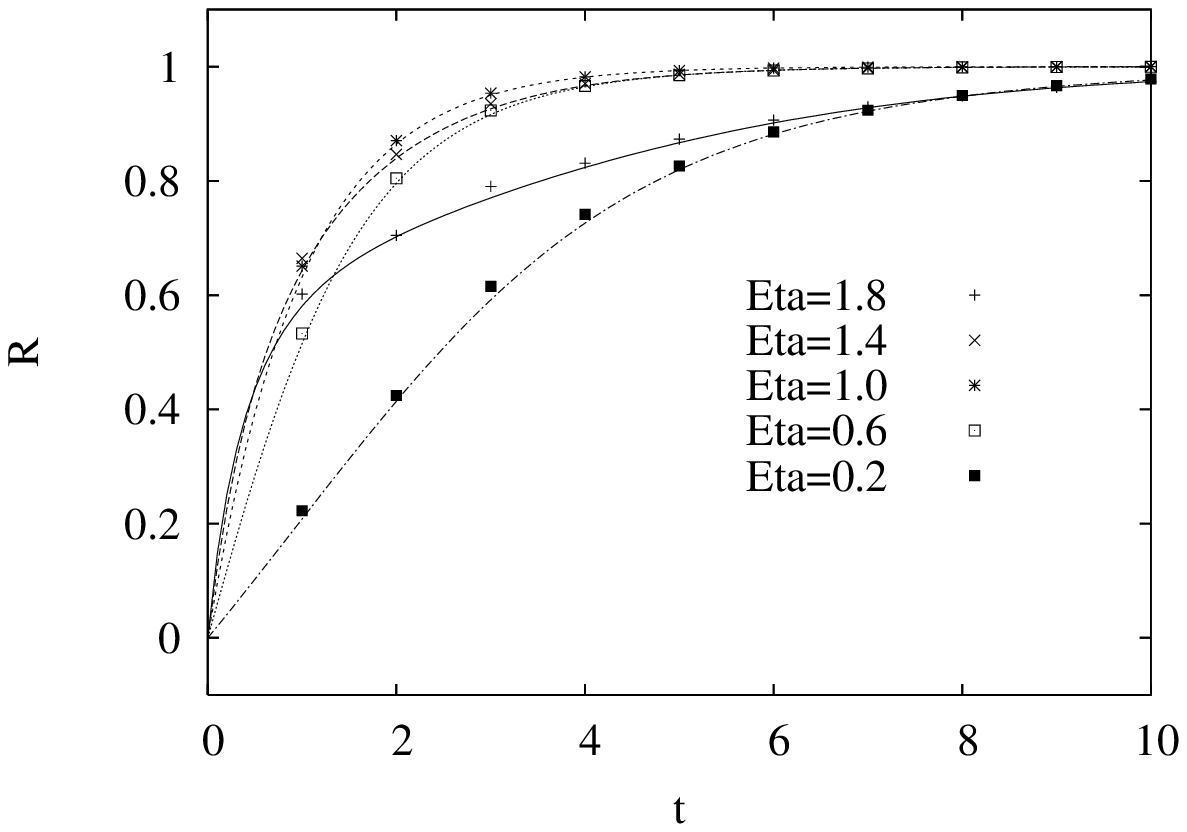}\\
(b) Dynamical behaviors of $R^t$.
\end{center}
\end{minipage}
%-----------------------------------------------
\caption{Dynamical behaviors of $l^t$ and $R^t$.
Theory and computer simulations.
$\sigma_B^2=\sigma_J^2=0.0$．}
\label{fig:lRallS00}
%-----------------------------------------------
\end{figure}

\begin{figure}[htbp]
\begin{minipage}{.500\linewidth}
\begin{center}
\includegraphics[width=\gsize\linewidth,keepaspectratio]{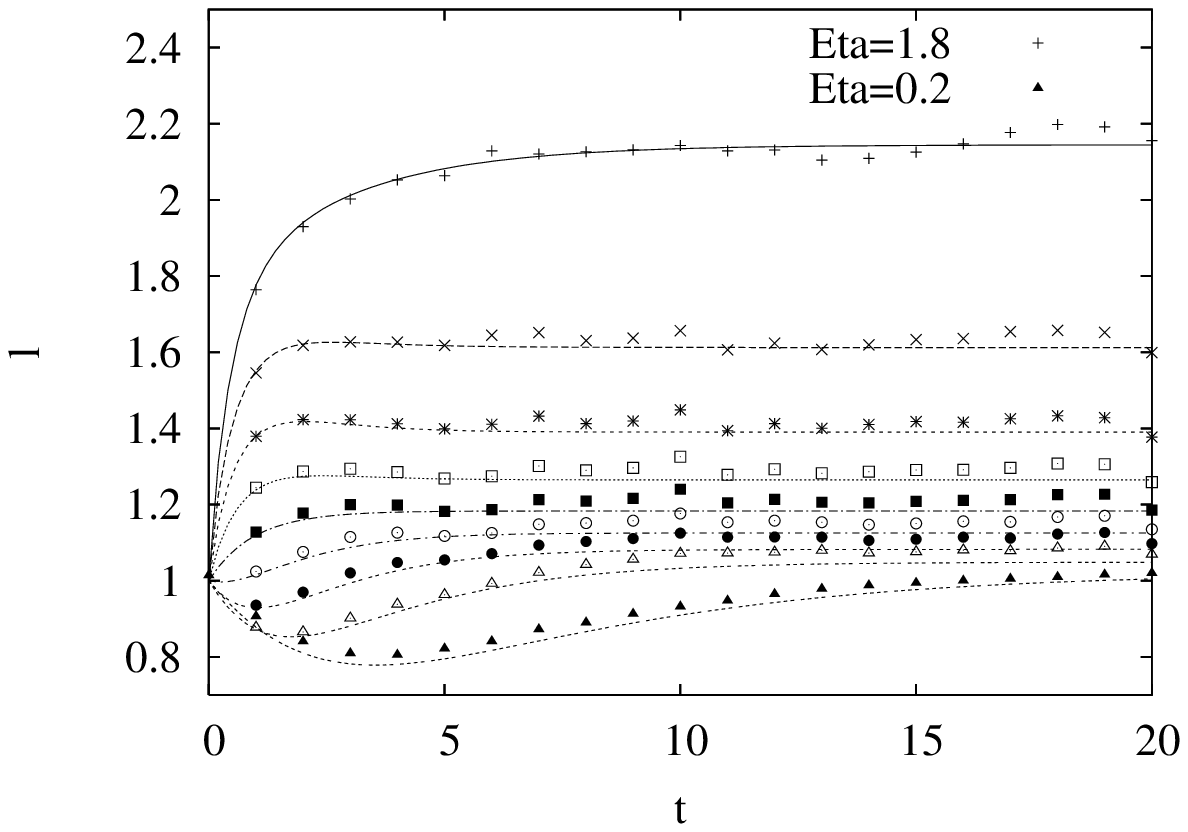}\\
(a) Dynamical behaviors of $l^t$.
$\eta=1.8, 1.6, 1.4, 1.2, 1.0, 0.8, 0.6, 0.4$ and $0.2$ from the top.
\label{fig:lall}
\end{center}
\end{minipage}
\begin{minipage}{.500\linewidth}
\begin{center}
\includegraphics[width=\gsize\linewidth,keepaspectratio]{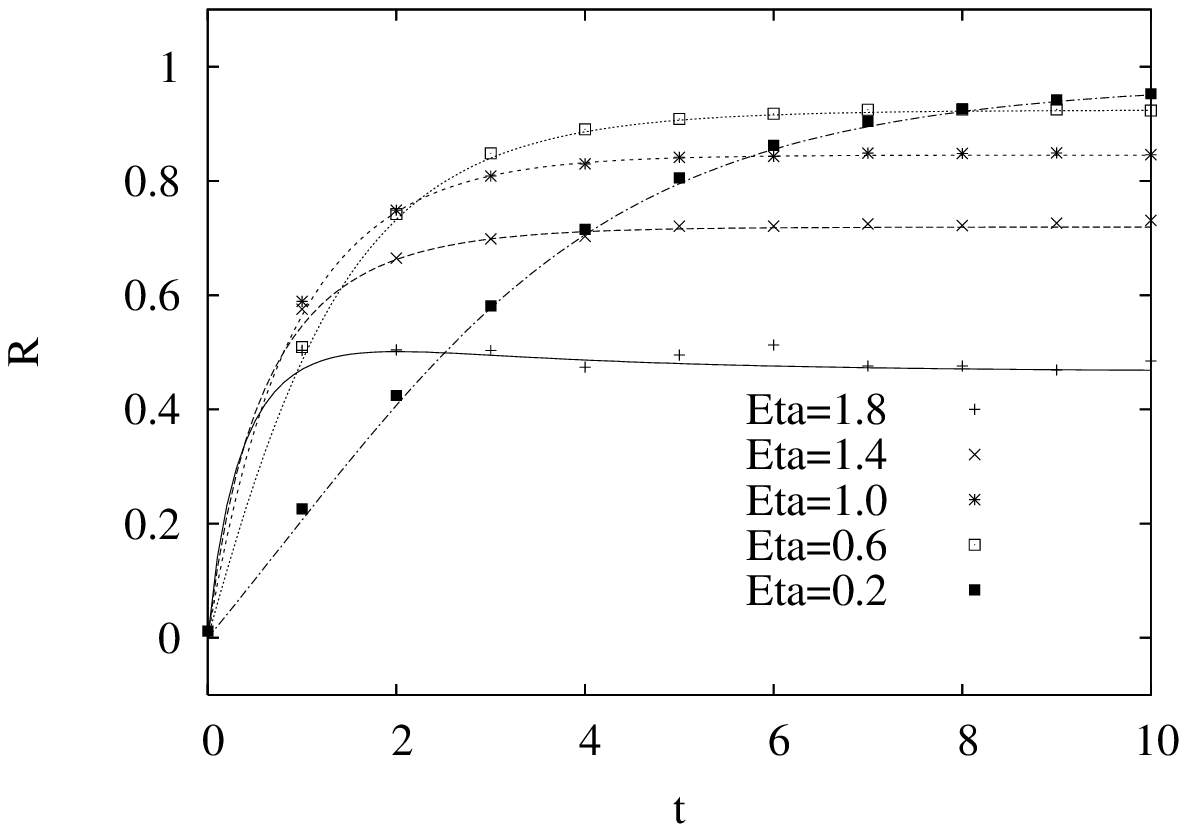}\\
(b) Dynamical behaviors of $R^t$.
\label{fig:Rall}
\end{center}
\end{minipage}
%-----------------------------------------------
\caption{Dynamical behaviors of $l^t$ and $R^t$.
Theory and computer simulations.
$\sigma_B^2=\sigma_J^2=0.0$．}
\label{fig:lRallS02}
%-----------------------------------------------
\end{figure}

%**********************************************************
The dynamical behaviors of $l^t$ and $R^t$
have been analytically obtained using 
eqs.(\ref{eqn:r}), (\ref{eqn:rsol}), and (\ref{eqn:l2sol}).
Figures \ref{fig:lRallS00} and \ref{fig:lRallS02}
show some examples of the analytical results 
and the corresponding
simulation results, where $N=2000$.
In these figures, 
the curves represent theoretical results. 
The symbols 
represent simulation results.
Figure \ref{fig:lRallS00} shows the results of 
$\sigma_B^2=\sigma_J^2=0.0$
and no noise.
Figure \ref{fig:lRallS02} shows the result of
$\sigma_B^2=\sigma_J^2=0.2$.

Focusing on the signs of the powers
of the exponential functions in eq.(\ref{eqn:l2sol}),
we can see that 
$l^t$ diverges if the learning rate satisfies
$0>\eta$ or $\eta>2$.
$l^t$ converges to 
\begin{equation}
%l^{\infty}=\sqrt{1+\frac{\eta}{2-\eta}\left(\sigma_B^2+\sigma_J^2\right)}
l^{\infty}=\sqrt{1+\bar{\sigma}^2}
\label{eqn:linfty}
\end{equation}
%\bar{\sigma}^2
if $0<\eta<2$.
Equations (\ref{eqn:r}) and (\ref{eqn:rsol})
show that $R^t$ converges to 
\begin{equation}
R^{\infty}=1/l^{\infty}.
\label{eqn:Rinfty}
\end{equation}
%$R^{\infty}=1/l^{\infty}$

Therefore, we can see that
$l^{\infty}=R^{\infty}=1$ in the case of no noise and
$l^{\infty}>1,\ R^{\infty}<1$ in the case of noise.

Since eq.(\ref{eqn:l2sol}) shows 
\begin{equation}
\left. \frac{d(l^t)^2}{dt}\right|_{t=0} \cases{
< 0 & when\ $\eta<\frac{2}{2+\sigma_B^2+\sigma_J^2}$, \cr
= 0 & when\ $\eta=\frac{2}{2+\sigma_B^2+\sigma_J^2}$, \cr
> 0 & when\ $\eta>\frac{2}{2+\sigma_B^2+\sigma_J^2}$, \cr
}
\label{eqn:ldif}
\end{equation}
an equation regarding $t$
\begin{eqnarray}
\frac{d(l^t)^2}{dt} &=& 0
\label{eqn:ldif0}
\end{eqnarray}
has only one solution
\begin{eqnarray}
t=\frac{1}{\eta(1-\eta)}
\ln \left(2-\frac{\eta}{2}\left(2+\sigma_B^2+\sigma_J^2\right)\right),
\label{eqn:ldif0sol}
\end{eqnarray}
if the learning rate satisfies 
$0<\eta< 4/(2+\sigma_B^2+\sigma_J^2)$ and $\eta\neq 1$.
Therefore, $l^t$ asymptotically approaches unity
after becoming larger than unity if $0<\eta<1$
and $l^t$ asymptotically approaches unity
after becoming smaller than unity if $1<\eta<2$
as shown in Figure \ref{fig:lRallS00}(a).

Equations (\ref{eqn:r}), (\ref{eqn:rsol}), and (\ref{eqn:l2sol})
show
\begin{eqnarray}
\left. \frac{dR^t}{dt}\right|_{t=0} &=& \eta.
\label{eqn:}
\end{eqnarray}
Therefore, the larger $\eta$ is, the faster $R$ rises
as shown in Figs. \ref{fig:lRallS00}(b) and \ref{fig:lRallS02}(b).
However,
eqs.(\ref{eqn:r}), (\ref{eqn:rsol}), (\ref{eqn:l2sol}), 
(\ref{eqn:linfty}), and (\ref{eqn:Rinfty}) show
\begin{eqnarray}
R^{\infty}-R^t &=& \frac{1}{l^{\infty}}-\frac{r^t}{l^t} \\
&\rightarrow& \left(1+\bar{\sigma}^2\right)
          \left(\bar{\sigma}^2 e^{-\eta t}
          +\frac{2-\bar{\sigma}^2}{2} e^{\eta(\eta-2)t} \right), 
\label{eqn:Rcnv}
\end{eqnarray}
when $t$ is large.
Since eq.(\ref{eqn:Rcnv}) 
is $O(e^{-\eta t})$ if $0<\eta\leq 1$
and $O\left(e^{\eta(\eta-2)t}\right)=
O\left(e^{((\eta-1)^2-1)t}\right)$ if $1<\eta <2$,
the convergence of $R^t$ is the fastest
when the learning rate satisfies $\eta=1$.
This can be confirmed 
in Figure \ref{fig:lRallS00}(b) and Figure \ref{fig:lRallS02}(b).
This phenomenon can be understood by the fact that
$\eta=1$ is a special condition
where the student uses up the information
obtained from input $\mbox{\boldmath $x$}$\cite{JPSJ2006b}.

\begin{figure}[htbp]
\begin{minipage}{.500\linewidth}
\begin{center}
\includegraphics[width=\gsize\linewidth,keepaspectratio]{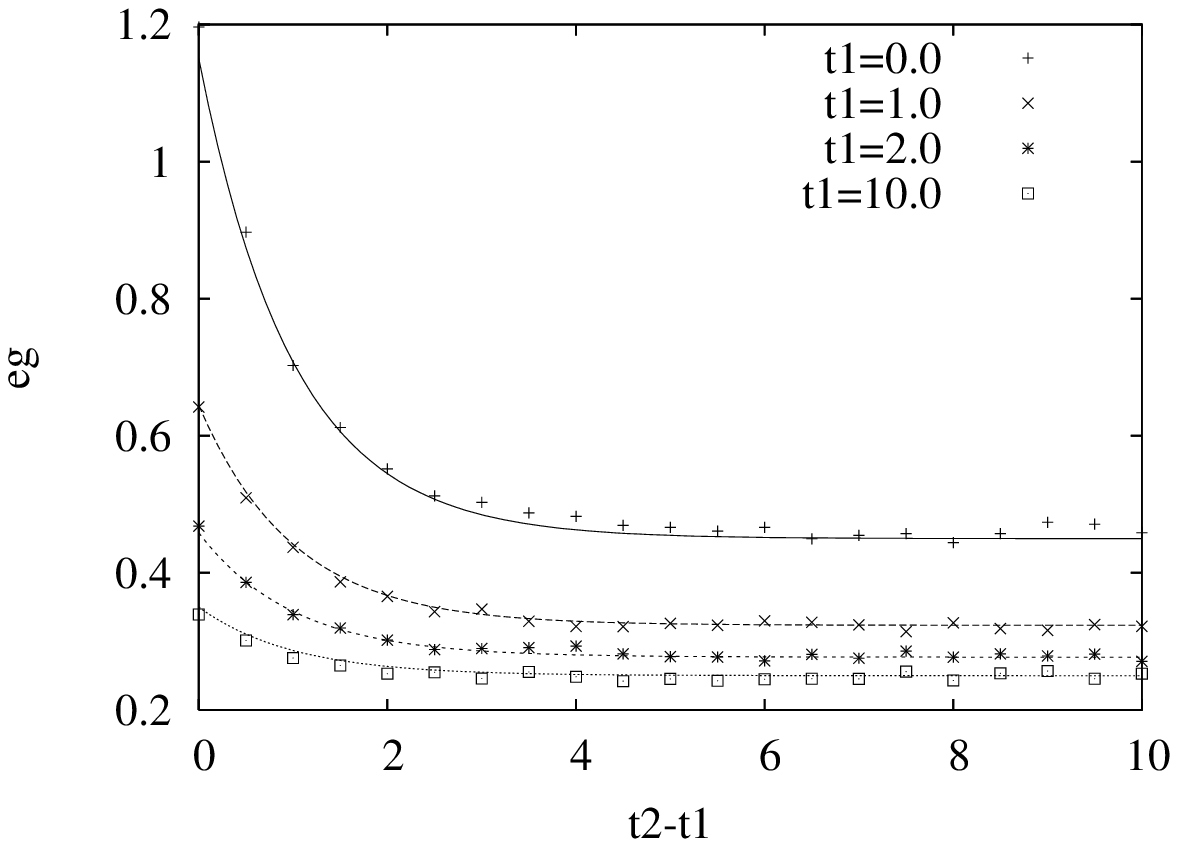}\\
(a) $\epsilon_g$
\end{center}
\end{minipage}
\begin{minipage}{.500\linewidth}
\begin{center}
\includegraphics[width=\gsize\linewidth,keepaspectratio]{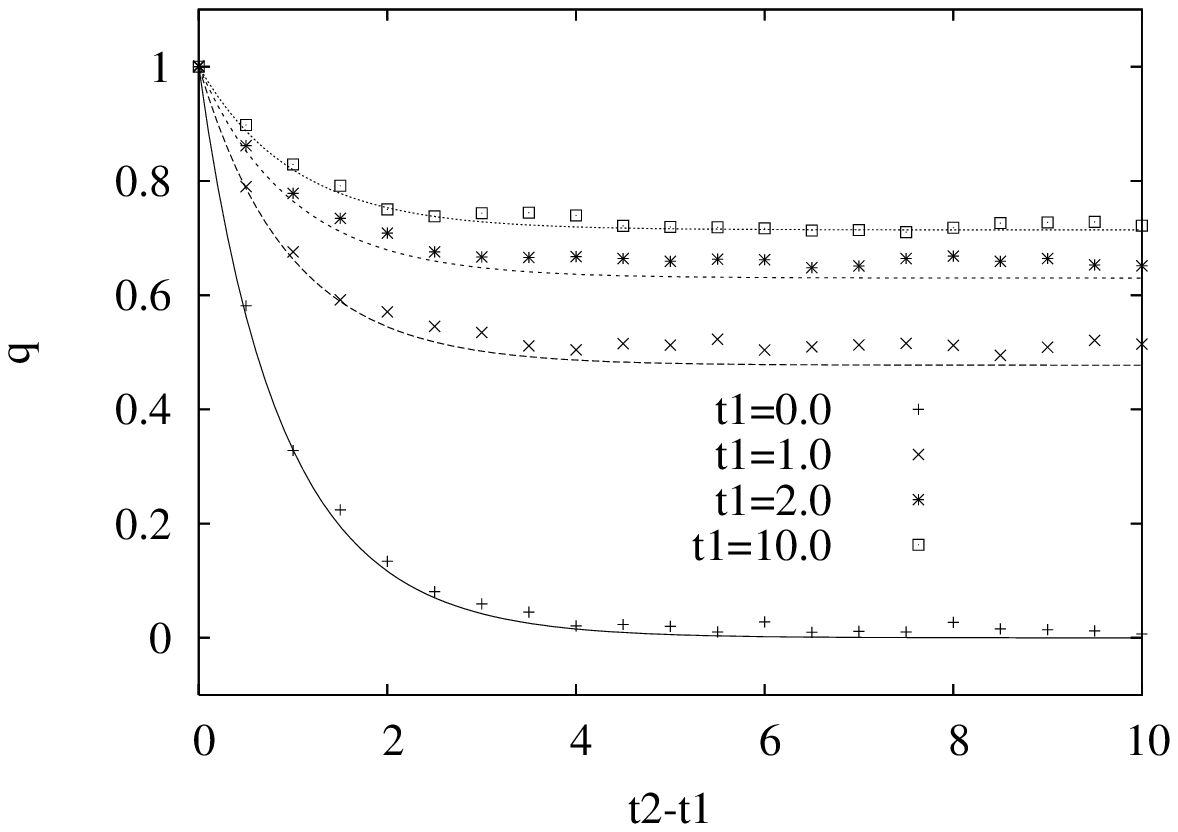}\\
(b) $q^{t_1,t_2}$
\end{center}
\end{minipage}
%-----------------------------------------------
\caption{Relationship between 
$t_2-t_1$ and $\epsilon_g$, $q^{t_1,t_2}$
in the case of constant leading time $t_1$.
Theory and computer simulations.
$\eta=1.0, \sigma_B^2=\sigma_J^2=0.2$．}
\label{fig:eg1S02}
%-----------------------------------------------
\end{figure}

\begin{figure}[htbp]
\begin{minipage}{.500\linewidth}
\begin{center}
\includegraphics[width=\gsize\linewidth,keepaspectratio]{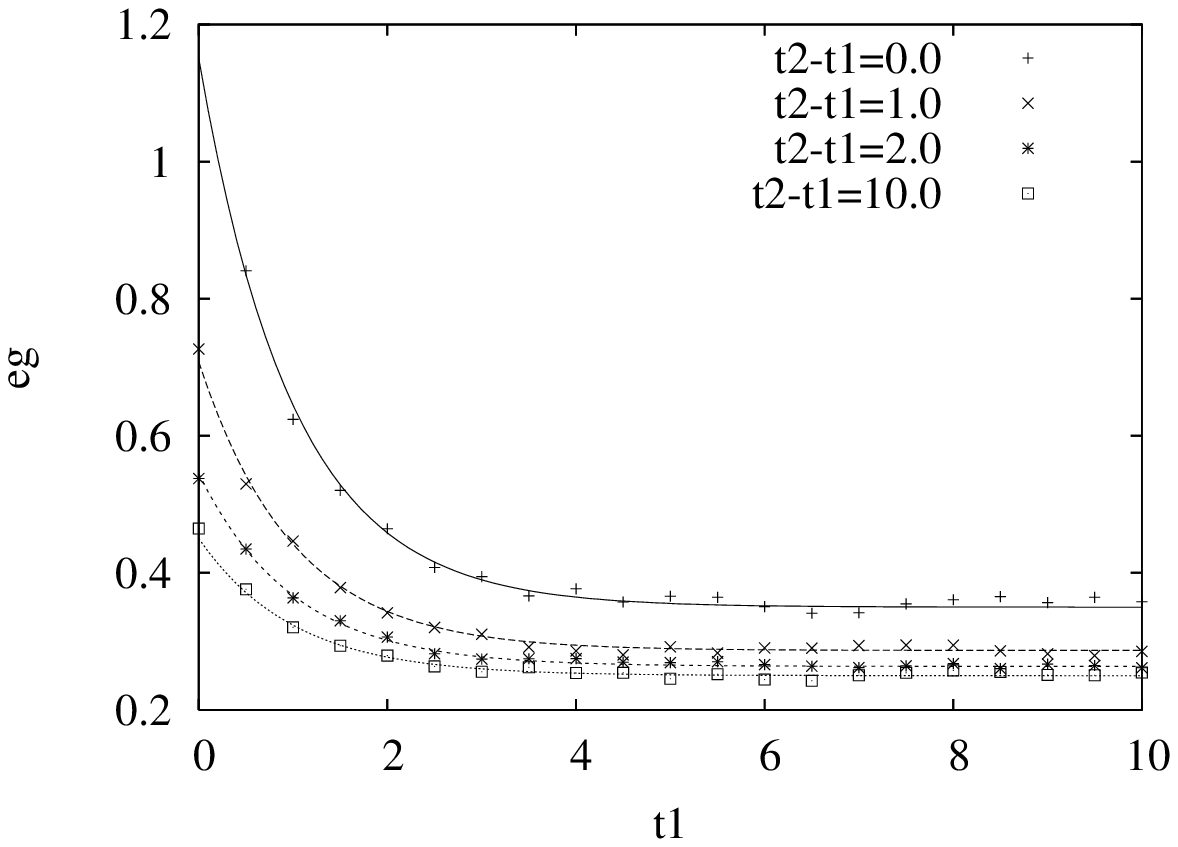}\\
(a) $\epsilon_g$
%-----------------------------------------------
%-----------------------------------------------
%\label{fig:eg1ETA10}
\end{center}
\end{minipage}
\begin{minipage}{.500\linewidth}
\begin{center}
\includegraphics[width=\gsize\linewidth,keepaspectratio]{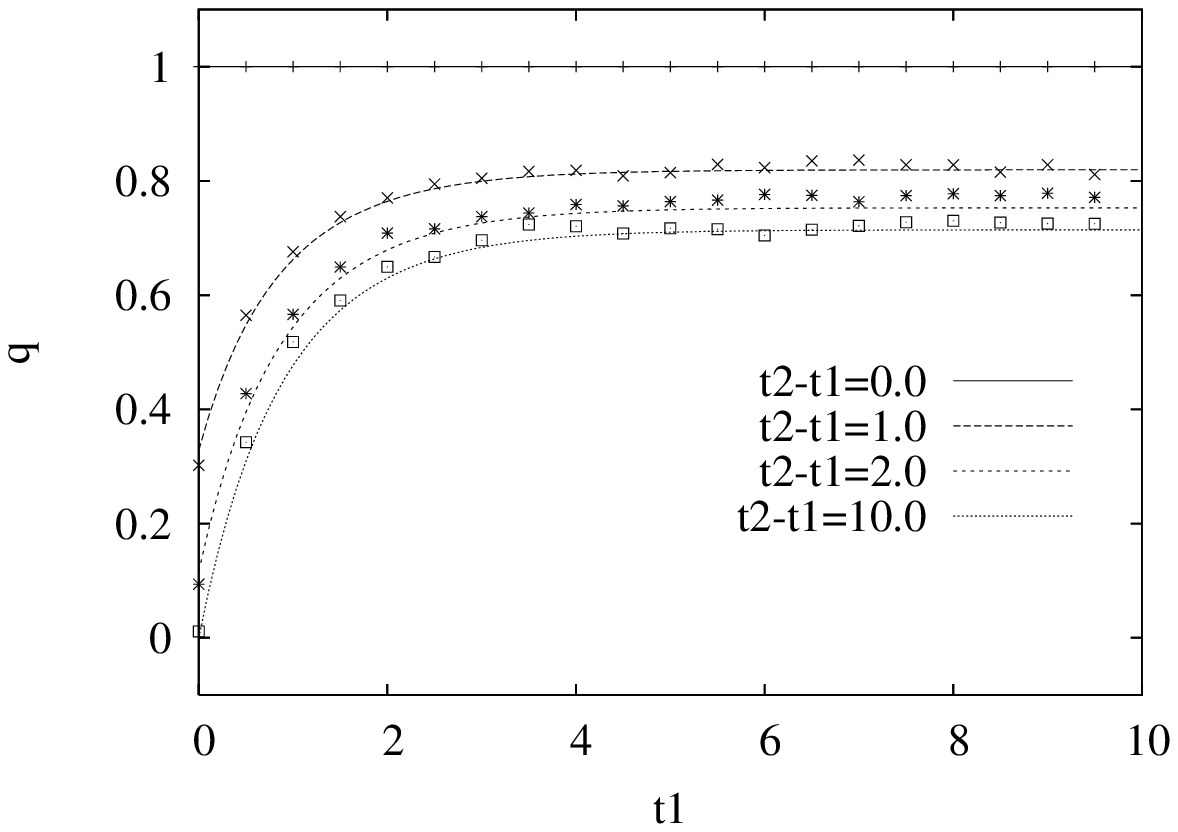}\\
(b) $q^{t_1,t_2}$
%-----------------------------------------------
%-----------------------------------------------
%\label{fig:q1ETA10}
\end{center}
\end{minipage}
%-----------------------------------------------
\caption{Relationship between 
$t_1$ and $\epsilon_g$, $q^{t_1,t_2}$
in the case of constant time interval $t_2-t_1$.
Theory and computer simulations.
$\eta=1.0, \sigma_B^2=\sigma_J^2=0.2$．}
\label{fig:eg2S02}
%-----------------------------------------------
\end{figure}

We analytically obtained the dynamical behaviors of the generalization error 
$\epsilon_g$ and the direction cosine $q$
using 
eqs.(\ref{eqn:Q}) and (\ref{eqn:l2sol})--(\ref{eqn:sbar}).
Figures \ref{fig:eg1S02} and \ref{fig:eg2S02}
show some examples of the analytical results 
and the corresponding
simulation results, where $N=2000$.
In these figures, 
the curves represent theoretical results. 
The symbols 
represent simulation results.
$\epsilon_g$ has been calculated 
for the simplest case, that is $K=2, C_1=C_2=1/2$.
Other conditions are 
$\eta=1.0$ and $\sigma_B^2=\sigma_J^2=0.2$.
In the computer simulations, 
$\epsilon_{g}$
was obtained by
averaging the squared errors for $10^4$ random inputs 
at each time step.

Figure \ref{fig:eg1S02} shows the relationship 
between $t_2-t_1$ and $\epsilon_g$, $q^{t_1,t_2}$
in the case of a constant $t_1$.
%$\epsilon_g$ decreases monotonically, remains constant, or
%increases monotonically
%when $t_2-t_1$ increases.
%The reason this phenomenon can be explained is as follows:
When $t_2-t_1$ increases, 
$\epsilon_g$ increases monotonically, remains constant, or
decreases monotonically depending on the values of $\eta$.
We prove this in the following.
Equation (\ref{eqn:eg2}) shows that
$\epsilon_{g(K=1)}$ is
\begin{eqnarray}
\epsilon_{g(K=1)} &=& \frac{1}{2}
\left(\frac{\eta}{2-\eta}\left(\sigma_B^2+\sigma_J^2\right)
+
\left(2-\frac{\eta}{2-\eta}\left(\sigma_B^2+\sigma_J^2\right)\right)
e^{\eta(\eta-2)t}\right).
\label{eqn:egK1}
\end{eqnarray}
Therefore, $\epsilon_{g(K=1)}$
decreases monotonically, remains constant, or
increases monotonically as time $t$ proceeds.
The necessary and sufficient conditions 
for the above three phenomena are
\begin{eqnarray}
\eta &<& \frac{4}{2+\sigma_B^2+\sigma_J^2}, \label{eqn:egdec}\\
\eta &=& \frac{4}{2+\sigma_B^2+\sigma_J^2}, \label{eqn:egconst}\\
\eta &>& \frac{4}{2+\sigma_B^2+\sigma_J^2}, \label{eqn:eginc}
\end{eqnarray}
respectively.
Since the output of the ensemble 
is the weighted sum of the outputs of the brothers,
the generalization error for $K>1$ also
decreases monotonically, remains constant, or
increases monotonically.
The necessary and sufficient conditions 
for these three phenomena are also
shown in eqs.(\ref{eqn:egdec})--(\ref{eqn:eginc}).
Since the condition of Fig. \ref{fig:eg1S02}(a)
agrees with eq.(\ref{eqn:egdec}), the generalization error
decreases monotonically.
Equations (\ref{eqn:Q}), (\ref{eqn:l2sol}), (\ref{eqn:Qsol}),
and (\ref{eqn:linfty}) show
that
$q^{t_1,t_2}$ in the case of $t_1=0$
asymptotically approaches zero
when $t_2-t_1\rightarrow \infty$ as shown 
in Fig. \ref{fig:eg1S02}(b).
This means that after a long time the student
is orthogonal with its initial condition.

Since the order parameters and $\epsilon_g$
have been explicitly obtained as functions of $t$ and $t'$
as eqs.(\ref{eqn:rsol})--(\ref{eqn:eg2})
in this paper,
the relationships between $t_1$ and $\epsilon_g$, $q^{t,t'}$
in the case of constant time interval of the brothers 
or
constant $t_{k+1}-t_k$ can be calculated.
Figure \ref{fig:eg2S02} shows the relationship 
between $t_1$ and $\epsilon_g$, $q^{t_1,t_2}$
in the case of constant $t_2-t_1$.
For the same reason as in Fig. \ref{fig:eg1S02}(a),
the generalization error $\epsilon_g$ also decreases 
monotonically in Fig. \ref{fig:eg2S02}(a).
Figure \ref{fig:eg2S02}(b) shows that
$q^{t_1,t_2}$ converges to a smaller value than 
unity in the case of $t_2-t_1\neq 0.0$.
This means that the student itself continues to move
after the order parameters $l, R$, and $q$
reach a steady state.

In Figs. \ref{fig:eg1S02} and \ref{fig:eg2S02},
the generalization error $\epsilon_g$
and the direction cosine $q^{t_1,t_2}$
seem to reach a almost steady state
by $t_2-t_1>5$ or $t_1>5$.
The behaviors of $\epsilon_g$ when 
the leading time $t_1 \rightarrow \infty$
or 
the time interval $t_{k+1}-t_k \rightarrow \infty$
can be theoretically obtained 
since the generalization error and the 
order parameters have been analytically obtained 
as functions of $t_k,\ k=1,\ldots,K$,
as shown in eq.(\ref{eqn:eg2}).

At first, eqs.(\ref{eqn:l2sol}) and (\ref{eqn:eg2})
show that $(l^t)^2$ diverges unless $0<\eta<2$.
Therefore, the generalization error diverges 
unless $0<\eta<2$.
If $0<\eta<2$, the generalization error
can be discussed as follows:

When $t_1 \rightarrow \infty$, 
from eqs.(\ref{eqn:eg2}) and (\ref{eqn:sbar}),
we obtain
\begin{eqnarray}
\epsilon_g
&=& \frac{1}{2}\left[1-2\sum_{k=1}^K C_k 
+ 2\sum_{k=1}^K \sum_{k'>k}^K C_k C_{k'}
\left(1+\frac{\eta}{2-\eta}\left(\sigma_B^2+\sigma_J^2\right)
e^{-\eta (t_{k'}-t_k)} \right) \right. \nonumber \\
&+& \left. \sum_{k=1}^K C_k^2
\left(1+\frac{\eta}{2-\eta}\left(\sigma_B^2+\sigma_J^2\right)\right)
+\sigma_B^2+\sum_{k=1}^K C_k^2 \sigma_J^2
\right].
\label{eqn:eg3}
\end{eqnarray}

In addition, when the time interval 
$t_{k+1}-t_k \rightarrow \infty$,
from eq.(\ref{eqn:eg3}),
we obtain
\begin{eqnarray}
\epsilon_g
&=& \frac{1}{2}\left[1-2\sum_{k=1}^K C_k
+ 2\sum_{k=1}^K \sum_{k'>k}^K C_k C_{k'}
+ \sum_{k=1}^K C_k^2
\left(1+\frac{\eta}{2-\eta}\left(\sigma_B^2+\sigma_J^2\right)\right)
+\sigma_B^2+\sum_{k=1}^K C_k^2 \sigma_J^2
\right].
\label{eqn:eg4}
\end{eqnarray}

Equation (\ref{eqn:eg4}) shows
that the generalization error decreases
as the learning rate $\eta$ decreases
regardless of $K$
when $t_1 \rightarrow \infty$ and 
$t_{k+1}-t_k \rightarrow \infty$.

In addition,
when the weights are uniform or $C_k=C=1/K$,
from eq.(\ref{eqn:eg4}), we obtain
\begin{eqnarray}
\epsilon_g
&=& \frac{1}{2K}
\left(\frac{\eta}{2-\eta}\left(\sigma_B^2+\sigma_J^2\right)\right)
+\frac{1}{2}\left(\sigma_B^2 + \frac{1}{K}\sigma_J^2\right).
\label{eqn:eg6}
\end{eqnarray}

Here, considering the special case $K=1$, we obtain
\begin{eqnarray}
\epsilon_g
&=& \frac{1}{2}
\left(\frac{\eta}{2-\eta}\left(\sigma_B^2+\sigma_J^2\right)\right)
+\frac{1}{2}\left(\sigma_B^2 + \sigma_J^2\right).
\label{eqn:eg63}
\end{eqnarray}
If 
$\mbox{\boldmath $B$}=\mbox{\boldmath $J$}^{t_1}$
is true, 
the generalization error must equal
the residual error 
\begin{eqnarray}
\epsilon_g
&=& \frac{1}{2}\left(\sigma_B^2 + \sigma_J^2\right)
\label{eqn:eg62}
\end{eqnarray}
caused by noise
from eq.(\ref{eqn:e}), which is the definition of error.
Therefore, the difference between eq.(\ref{eqn:eg63})
and eq.(\ref{eqn:eg62})
\begin{equation}
\frac{1}{2}
\left(\frac{\eta}{2-\eta}\left(\sigma_B^2+\sigma_J^2\right)\right)
\end{equation}
is caused by the disagreement
between $\mbox{\boldmath $B$}$
and $\mbox{\boldmath $J$}^{t_1}$.

Next, let us consider another special case,
$K=\infty$.
If and only if
\begin{equation}
\mbox{\boldmath $B$}=\lim_{K\rightarrow \infty}
\frac{1}{K}\sum_{k=1}^K \mbox{\boldmath $J$}^{t_k},
\label{eqn:BKinfty}
\end{equation}
the generalization error must equal
the residual error 
\begin{eqnarray}
\epsilon_g
&=& \frac{1}{2}\sigma_B^2
\label{eqn:eg65}
\end{eqnarray}
caused by noise
from eq.(\ref{eqn:e}), which is the definition of error.
Equation (\ref{eqn:BKinfty}) is true
since eq.(\ref{eqn:eg6}) equals eq.(\ref{eqn:eg65})
when $K=\infty$.

In addition, if $\sigma_B^2=\sigma_J^2=\sigma^2$,
eq.(\ref{eqn:eg6}) changes as follows:
\begin{eqnarray}
\epsilon_g
&=& \left(\frac{1}{2K}\frac{2+\eta}{2-\eta}+\frac{1}{2}\right)\sigma^2.
\label{eqn:eg7}
\end{eqnarray}

\begin{figure}[htbp]
\begin{center}
\includegraphics[width=0.600\linewidth,keepaspectratio]{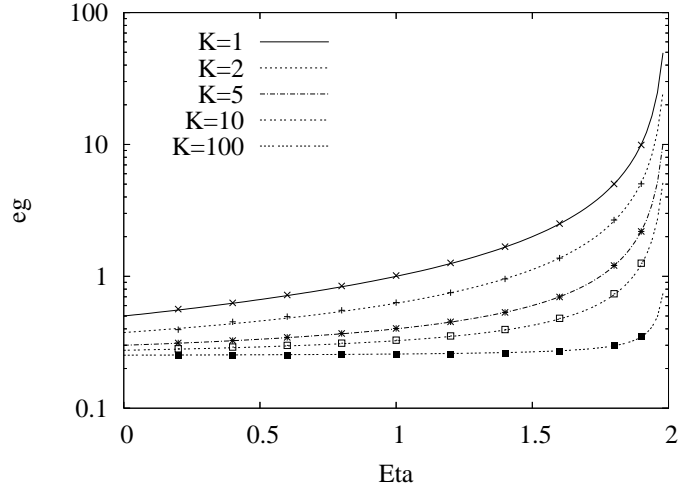}
%-----------------------------------------------
\caption{
Relationship between the learning rate $\eta$
and the generalization error $\epsilon_g$,
when both the leading time $t_1$ and the time interval
$t_{k+1}-t_k$ are large enough. 
Theory and computer simulations.
Conditions are $C_k=C=1/K$ and $\sigma_B^2=\sigma_J^2=0.5$.
}
%-----------------------------------------------
\label{fig:eg-eta}
\end{center}
\end{figure}

The relationship between the learning rate $\eta$
and the generalization error $\epsilon_g$
can be analytically obtained using 
eq.(\ref{eqn:eg7})
when both the leading time $t_1$ and the time interval
$t_{k+1}-t_k$ are large enough,
and the uniform weight $C_k=C=1/K$ and $\sigma_B^2=\sigma_J^2=0.5$.
Figure \ref{fig:eg-eta}
shows the analytical results 
and the corresponding
simulation results.
In the computer simulations, $N=2000$,
the leading time $t_1=10$, and the time interval $t_{k+1}-t_k=10$
($t_1=t_{k+1}-t_k=20$ when $\eta=0.2$),
we obtained $\epsilon_{g}$
by
averaging the squared errors for $10^4$ random inputs 
at each time step.
Figure \ref{fig:eg-eta} confirms the following.
The generalization error decreases as
the learning rate $\eta$ decreases.
The generalization error decreases 
and converges to the residual error $\frac{1}{2}\sigma_B^2$
as $K$ increases.

In addition, if the learning rate satisfies $\eta=1$,
eq.(\ref{eqn:eg7}) becomes
\begin{eqnarray}
\epsilon_g
&=& \left(\frac{3}{2K}+\frac{1}{2}\right)\sigma^2.
\label{eqn:eg8}
\end{eqnarray}

Equation (\ref{eqn:eg8}) refers to 
the generalization error $\epsilon_g$ of $K=\infty$,
which is $1/4$ times of that of $K=1$ when 
the learning rate satisfies $\eta=1$, 
the uniform weights $C_k=1/K$,
$\sigma_B^2=\sigma_J^2$, 
$t_1 \rightarrow \infty$, and $t_{k'}-t_k \rightarrow \infty$.
%Comparing this with the fact
%that the ratio is $1/2$ of the conventional space domain
%ensemble learning with $\eta=1$, $C_k=1/K$ and 
%$\sigma_B^2=\sigma_J^2$,
Since the generalizaion error $\epsilon_g$ of 
the conventional 
space domain ensemble learning with 
$K=\infty$, $\eta=1$, $C_k=1/K$ and 
$\sigma_B^2=\sigma_J^2$
is $1/2$ times of that of $K=1$\cite{Hara}, 
we can say that the time domain
ensemble learning is twice as effective as 
the conventional space domain ensemble learning.
We can explain this difference as follows:
In conventional space domain
ensemble learning, the similarities among students
become high since all students use the same examples for learning.
On the other hand, in time domain ensemble learning,
the similarities among brothers become low
since all brothers use almost totally different examples for learning.

%**********************************************************
\section{Conclusion}
%**********************************************************
We analyzed the generalization performance
of the time domain ensemble learning
in the framework of online learning 
using a statistical mechanical method.
We treated a model in which
both the teacher and the student
were linear perceptrons with noises.
We showed that 
the time domain
ensemble learning is twice as effective as
the conventional space domain
ensemble learning.

%**********************************************************
\section*{Acknowledgments}
%**********************************************************
%\noindent
%{\bf 謝辞} 
This research was partially supported by the Ministry of Education, 
Culture, Sports, Science, and Technology of Japan, 
with Grants-in-Aid for Scientific Research
14084212, 15500151 and 16500093.
%本論文の一部は科学研究費補助金　
%特定領域研究
%（課題番号
%岡田 ベイズ推定に基づくデコード過程に関する統計力学的理論とその応用
%特定領域研究　
%14084212），
%同　基盤研究(C)
%（課題番号
%岡田 画像修復の知見を用いたガウス過程関数近似の統計的理論とその応用
%基盤研究(C)　
%14580438, 
%三好 移動する教師に対するオンライン学習の解析
%基盤研究(C)
%15500151, 
%岡田 相転移現象を利用した離散・連続混在型情報圧縮アルゴリズムの研究
%基盤研究(C)
%16500093
%）に
%よるものであり，ここに感謝いたします．

%**********************************************************

\end{document}